\documentclass[aps,prd,superscriptaddress,twocolumn, floatfix]{revtex4-1}

\usepackage[utf8]{inputenc}
\usepackage{tikz}
\usetikzlibrary{shapes.geometric}
\usepackage{xcolor}
\usepackage{placeins}
\usepackage{verbatim}
\usepackage{graphicx}
\usepackage{dcolumn}
\usepackage{bm}
\usepackage{hyperref}
\usepackage{float}
\usepackage{amssymb,amsmath}
\usepackage{amsthm}
\usepackage[english]{babel}
\DeclareMathOperator{\sech}{sech}
\begin{document}
\title{Metamaterial branes}
\author{F. A. P. Alves-J\'{u}nior}
\email{francisco.artur@univasf.edu.br} 
\affiliation{Universidade Federal do Vale do S\~{a}o Francisco, \textit{Campus} Serra da Capivara, Brazil}

\author{A. B. Barreto}
\email{adriano.barreto@caxias.ifrs.edu.br}
\affiliation{Instituto Federal de Educa\c{c}\~{a}o, Ci\^{e}ncia e Tecnologia do Rio Grande do Sul, \textit{Campus} Caxias do Sul, Brazil}
\affiliation{Departamento de F\'{i}sica, Universidade Federal de Santa Catarina, Florian\'{o}polis, Brazil}

\author{F. Moraes}
\email{fernando.jsmoraes@ufrpe.br}
\affiliation{Departamento de F\'{i}sica, Universidade Federal Rural de Pernambuco, 52171-900 Recife, PE, Brazil}
 

\date{\today}
\begin{abstract}
In this article, we propose metamaterials analog brane models based on the geometric optics approach.
We show how to model the Randall-Sundrum thin brane model and the Gremm model for a thick brane.  We incorporate the Yukawa-like confinement mechanics for thick branes in the analog system and show an asymmetrical analog brane model with and without the confinement mechanisms. 
\end{abstract}

\maketitle
\section{Introduction\label{sec:Intro}}
The extra-dimensional theories were first guided by the unification of gravitational and electromagnetic interactions, this is the case of the Kaluza-Klein theories\cite{wesson2006five}. Brane-world models, on the other hand, are motivated by hierarchy problem \cite{arkani1998hierarchy,mannheim2005brane} and the cosmological constant problem \cite{weinberg1989cosmological}, besides other important questions in particle physics \cite{fitzpatrick2008flavor,feruglio2004extra}. In particular, the Randall-Sundrum models \cite{randall1999alternative,randall1999large} assume that the universe is a five-dimensional semi-Riemannian manifold, with a particular four-dimensional slice, the so-called brane, in which gravity is the weakest fundamental interaction. Furthermore, these models contain another important hypothesis, that the universe is described by warped geometry \cite{Felder:2001da}. 



According to brane-world models, the fifth dimension is only accessible for high energy phenomena, as a result, many proposals are trying to find its footprints by using the particle detectors \cite{kretzschmar2016searches}. At the quantum regime such as the proton radius problem seems to be evidence of the extra-dimensions \cite{dahia2016proton,wang2013proton} and the atomic spectroscopy seems to give extra-dimensional signs \cite{dahia2016constraints,lemos2019spectroscopic}.
On the other hand, analog gravity models have given many insights into curved space-time physics,  as  in the analog black hole old and new experiments \cite{jacobson1991black,drori2019observation,barcelo2011analogue} that shed some light on Hawking radiation. Particularly, metamaterials have appeared in many papers to simulate gravitational effects such as the Milne cosmology \cite{figueiredo2016modeling},  Klein analog space-times \cite{alves2021implications,smolyaninov2010metric},  black holes \cite{fernandez2016anisotropic}, wormholes \cite{azevedo2021optical} and analog  cosmic strings \cite{fernandez2018wave}, for example. 

In a few words, metamaterials are artificial materials, first idealized by Veselago, about fifty-four years ago. He proposed a substance with negative permittivity ($\epsilon$) and permeability ($\mu$). This kind of matter does not naturally occur in nature, but nowadays using nanotechnology it is possible to construct sub-wavelength structures that effectively have $\epsilon$ and $\mu$ with great liberty, which means that we can have materials that are conductors in one direction and insulators on the other. In this way, many novel devices could be designed as the invisibility cloak \cite{leonhardt2009transformation}, very high-quality optical cavities \cite{Rho:13}, and transformation-optic objects \cite{chen2010transformation, Quach:13}.

Concerning extra-dimensional scenarios, it is known that metamaterials could reproduce some aspects of the Kaluza-Klein model (see for instance Ref.\cite{smolyaninov2010metamaterial}). 
In a complementary way, here, we propose a new optical analog space-time that mimics geodesics in warped geometries based on hyperbolic metamaterials.  To illustrate the versatility of our proposal, we also incorporate the Yukawa-like interaction in some examples, that is a modification in the line element, $ds^2$, due to brane scalar field,  proposed by \cite{dahia2007confinement} in the context of classical particle confinement problem as pointed out by \cite{la2008classical,romero2011fermion,guha2010brane,ghosh2010confinement}.

 This manuscript is organized as follows. In section I, we revisit some mathematical aspects of geodesics in warp geometries. In section II, we show two  metamaterial analog brane models, for thick and thin branes. We also apply the Yukawa-like interaction for classical particles in our model. In section III, we illustrate how metamaterials could realize asymmetrical branes, we analyze the scenarios with and without the scalar field interaction. In the last section, we close the paper with our final considerations.

\section{Warped geometry}

Previously, Ref. \cite{dahia2007warped,dahia2008geodesic}, it had been analyzed the classical confinement conditions arising from warped geometry. Given a line element
\begin{equation}
    ds^2=g_{ab}dx^adx^b,
\end{equation}
that describes a free-falling particle moving in a five-dimensional manifold, which has the affine geodesics given by
\begin{equation}
    \frac{d^2 x^a}{d\lambda^2}+\Gamma^{a }_{bc}\frac{d x^b}{d\lambda}\frac{d x^c}{d\lambda}=0,
\end{equation}
In the case of the following line element 
\begin{equation}\label{eq:RS-metric}
    ds^2 = e^{2f(w)} \eta_{\mu\nu}dx^{\mu}dx^{\nu} - dw^2,
\end{equation}
where $\eta^{\mu\nu}=\textrm{diag}(+1,-1,-1,-1)$ is the Minkowski metric, it is explicitly obtained the equation of a massive particle motion 
\begin{equation}\label{eq:original-equation}
    \frac{d^2 w}{d\lambda^2}+f'\left[1+\left( \frac{d w}{d \lambda}\right)^2\right]=0,
\end{equation}
the other directions:
\begin{equation}\label{brane-motion1}
    \frac{d^2x}{d\lambda^2}+f'\frac{dx}{d\lambda}\frac{d w}{d\lambda}=0,
\end{equation}
\begin{equation}\label{brane-motion2}
\frac{d^2y}{d\lambda^2}+f'\frac{dy}{d\lambda}\frac{dw}{d\lambda}=0,
\end{equation}
where the $f'$ denotes the derivative with respect to $w$. In this case, a set of solutions of \eqref{brane-motion1} and \eqref{brane-motion2} can be expressed as $x^{i}=x_0^{i}+v_{0}^{i}\int e^{-f(w)}d\lambda$, where $v_0^{i}$ is the three-dimensional initial velocity.  The solution  \eqref{eq:original-equation} of transverse motion depends on $f(w)$. However, we could use the relation 
\begin{equation}
   e^{2f} \left(\frac{dw}{d\lambda}\right)^2+(e^{-2f}-1)=\dot{w}_0^2 ,
   \label{EDO-w}
\end{equation}
to understand qualitatively or even solve the problem. This equation is in agreement with  \cite{dahia2007confinement} and is a modified conservation energy relation, where  $e^{2f}\left(\frac{dw}{d\lambda}\right)^2$ is a type of kinetic energy term and $e^{-2f}-1$ plays the role of an effective potential, $V_{eff}$.

In thick branes models \cite{dzhunushaliev2010thick}, the extra dimensions are infinite and beyond the warp factor $f(z)$, there is a scalar field $\phi$ in the description of this kind of structure. For the sake of the confinement of particles in these branes, there is an interesting proposal that uses a Yukawa-inspired coupling present in geodesic description \cite{dahia2007confinement}, that could  be easily written in the form
\begin{equation}
    ds^2 = e^{2F} \eta_{\mu\nu}dx^{\mu}dx^{\nu} + dw^2, 
\end{equation}
where F is a combination of the warp factor $f(z)$ and the scalar field $\phi$ given by
\begin{equation}\label{eq:new-warp}
    F=f+\frac{1}{2}\ln\left(m^2+h^2\phi^2\right).    
\end{equation}
Then the transverse motion of the particle is modified as
\begin{equation}
    \frac{d^2 w}{d\lambda^2}+F'\left[1+\left( \frac{d w}{d \lambda}\right)^2\right]=0,
\end{equation}
    and the  motion in other directions also is influenced by $\phi$. We can get the new equations by changing  $f\rightarrow F$ in \eqref{brane-motion1} and \eqref{brane-motion2}. 
    
To complete this section, we have to mention that the massless particle transverse motion is different from the other classical particles. The equation of photon along the extra dimension is
\begin{equation}\label{eq:photon}
     \frac{d^2 w}{d\lambda^2}+f'\left( \frac{d w}{d \lambda}\right)^2=0,
\end{equation}
and is the same equation for thin and thick branes, there is no coupling with the scalar field as in the case of massless particles.
\section{Warp geometry analogues}
Now we present a simple three-dimensional analog to warp geometry. Starting with the light propagating in a metamaterial, described by the analog metric 
\begin{equation}\label{eq:meta-metric}
    ds_{an}^2=\epsilon_{||}\left(dx^2+dy^2\right) +\epsilon_{\perp}dz^2.
\end{equation}
Let us set $z$ as the extra dimension in the analog system. Using the Fermat principle in order to study the light rays propagation in the metamaterial, we associate the geodesic equations
\begin{equation}
    \frac{d^2 x^{i}}{d\lambda^2}+\Gamma^{i}_{jk}\frac{d x^{j}}{d\lambda}\frac{dx^{k}}{d\lambda}=0,
\end{equation}
taking  $\lambda$ as the arc length parameter, the equation along $z$ direction may be expressed as
\begin{equation}
    \ddot{z}-\frac{1}{2\epsilon_{\perp}}\frac{d\ln \left[\epsilon_{||}\right]}{dz}\left(1-\epsilon_{\perp}\dot{z}^2 \right)=0 .
\end{equation}
So, we identify 
\begin{equation}
    f_{I}(z)=-\frac{1}{2\epsilon_{\perp}}\ln|\epsilon_{||}\left(z\right)|+c_0,
\end{equation}
where $c_0$ is an arbitrary constant. It is worth noting that the above metric maps not only \eqref{eq:RS-metric}, but also a class of metrics of the form $ds^2=e^{f(w)}\eta_{\mu\nu}dx^{\mu}dx^{\nu}\pm\alpha^2 dw^2$, where $\alpha$ is a non-zero real number and satisfies the condition $\epsilon_{\perp} = \mp\alpha^2$. As a consequence, the two-time extra-dimensions (Klein spacetimes) are also possible to realize, for example.

The motion in the $xy$-plane, the analog brane, is given by
\begin{equation}\label{eq:x-motion}
    \ddot{x}+\frac{d\ln \left[\epsilon_{||}\right]}{dz}\dot{z}\dot{x}=0 ,
\end{equation}
\begin{equation}\label{eq:y-motion}
    \ddot{y}+\frac{d\ln \left[\epsilon_{||}\right]}{dz}\dot{z}\dot{y}=0.
\end{equation}
To map the particle motion on the brane, we must choose
\begin{equation}
    f_{II}(z)=\frac{1}{2}\ln|\epsilon_{||}| + c_1.
\end{equation}
Setting $\epsilon_{\perp}=-1$ and $c_0=c_1=0$, we are able to find a general analog warp function $f(z)$ such that $f(z) = f_I(z) = f_{II}(z)$ and map the warped geometry into a metamaterial optical space. However, in order to mimic a thick brane geometry, it is convenient to use, as we shall see, the expression
   \begin{equation}\label{eq:F(z)}
       F(z)=\frac{1}{2}\ln\left(\epsilon_{||}\right),
   \end{equation}
where F(z) is given in \eqref{eq:new-warp}. This relation enables us to incorporate not only the metric in the permittivity tensor component but also the scalar field that smooths the brane geometry.

A curious feature of our model is that light particles in a metamaterial  mimic extra-dimensional massive particle geodesic motion in brane-world models. Usually, in metamaterial analog models, the analog mass is given by the electromagnetic wave frequency whereas, in our model, when taking into account
the Yukawa-like interaction, the analog mass is given by the tensor components $\epsilon_{||}$ or $\epsilon_{\perp}$, as we shall see in the models below. 

\subsection{Analogue thin brane geometry}
To mimic Randall-Sundrum  (RSI or RSII) geometries, we define a family of  hyperbolic metamaterials whose permeability tensor components are given by the expressions
\begin{equation}
    \epsilon_{||}=\exp\left ({ -2k|z|}\right)\quad \textrm{and} \quad \epsilon_{\perp}=-1.
\end{equation}

\begin{figure}
    \centering
    \includegraphics[scale = 0.6]{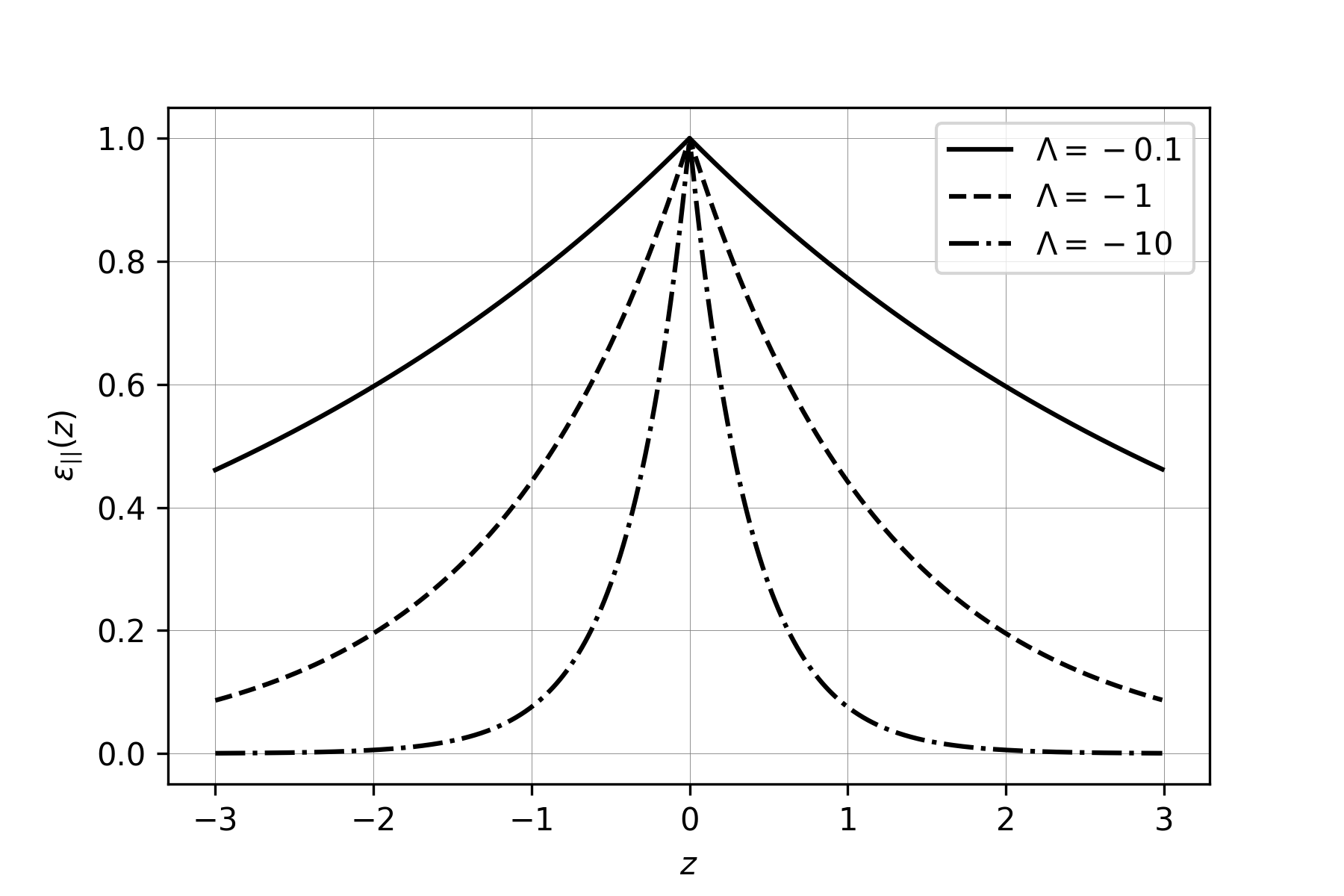}
    \caption{The profile of permittivity component $\epsilon_{||}$ against $z$, for different cosmological constants.}
    \label{fig:01}
\end{figure}
For every $k$ we have a particular type of anisotropic metamaterial.
We remember that $k$ is a parameter from the extra-dimensional model and is related to the anti-de Sitter five-dimensional cosmological constant $\Lambda$ through
\begin{equation}
    k=\pm\sqrt{-\frac{\Lambda}{6}},
\end{equation}
where $(+)$ stands for the original Randal--Sundrum models and $(-)$ for one of its variations \cite{muck2000geodesics}, here we deal just with the first case, but the other one is also possible. In both models, $k$ is related to the brane tension, $\sigma=-12k$.

In our model, for each cosmological constant value that we intend to mimic, we have to choose a particular $\epsilon_{||}$. See this relation in figure \eqref{fig:01}. To illustrate the particle behavior we use the dynamical equations present in the last section. Following \eqref{eq:x-motion} and \eqref{eq:y-motion}, the $xy-$plane  motion component is  
\begin{equation}\label{eq:meta-x-equation}
        \frac{dx}{d\lambda}=\frac{v_{0x}}{\epsilon_{||}(z_0)}\epsilon_{||},
\end{equation}

\begin{equation}\label{eq:meta-y-equation}
    \frac{dy}{d\lambda}=\frac{v_{0y}}{\epsilon_{||}(z_0)}\epsilon_{||}.
\end{equation}

In other words, the speeds $v_x$ and $v_y$ are influenced by $z-$coordinate. The transverse motion, which means the motion along the analog extra dimension, is 
\begin{equation}
    \begin{cases}
        \ddot{z} - k(\dot{z}^2+1) = 0, \, \textrm{for $z\geq 0$},\\
        \\
        \ddot{z} + k(\dot{z}^2+1) = 0, \, \textrm{for $z<0$}.
    \end{cases}
\end{equation}
It is easy to see that
\begin{equation}\label{eq:RS-analog}
    \frac{dz}{d\lambda}=\pm\textrm{tan}\left[k (\lambda-\lambda_0)\right],
\end{equation}
respectively for $z>0$ and  $z<0$. That leads to an attraction to the brane for $\lambda > \lambda_0$ and a repulsion from the brane for $\lambda < \lambda_0$. From this equation, for $\lambda\longrightarrow +\infty$, the particles get the speed of light. The solution of \eqref{eq:RS-analog} is given by
\begin{equation}
    z=z_0-\frac{1}{k}\ln{\left( \cos{k(\lambda -\lambda_0)}\right)}.
\end{equation}
\begin{figure}[h]
    \centering
    \includegraphics[scale = 0.6]{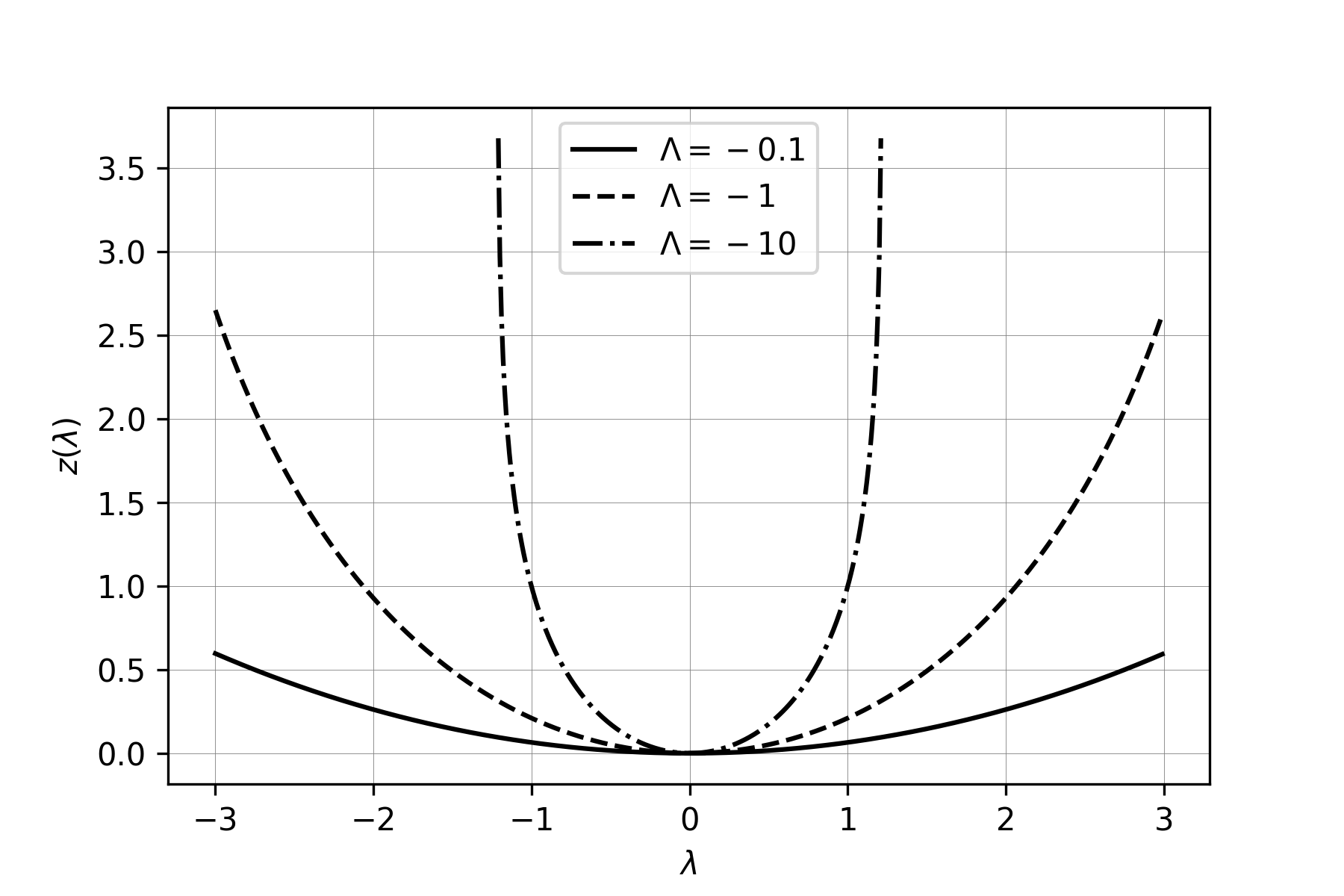}
    \caption{The motion of the analogue particles (light) taking $\lambda_0=0$ for different cosmological constants values.}
    \label{fig:02}
\end{figure}

Since the analog particles are photons, we could reproduce these initial conditions by sending a light ray perpendicular to $z-$direction. If we do that, it is remarkable that the light particles should be dragged or pulled in $z$-direction because of the metamaterial medium. From Fig. \eqref{fig:02}, we note that light is not confined at $z=0$, the analog brane region. Also, the particles at any initial position, $z_0> 0$, are attracted to the  origin, but once they reach it, they are repelled from there. Therefore, the plane $z=0$ behaves like a mirror. It is possible to see that sending particles from any $z_0<0$, also reflects in the brane region. As a result, particles from  one side could not migrate to the other side of the analog brane.

To compare these behaviors and the theoretical motion for free particles at extra dimensions found in \cite{rubakov2001large}, we point out the equation

\begin{equation}\label{eq:RS-original}
    w(t)=\frac{1}{2k}\ln\left(1+kt^2 \right),
\end{equation}
in terms of time-coordinate parameter. Rewriting, \eqref{eq:RS-original} in terms the proper-time parameter is possible to show that an equivalence between the behaviors of analogue particles and real massive particles at RS space-times.

\subsection{A analog model for thick brane geometry}
In thick brane models, we have the warp factor and we have one or more scalar fields defined along the extra-dimension that smooth the brane. Here we show how the thick domain wall could be mimicked with metamaterials. For this proposal, we follow  \cite{gremm2000four}, where the corresponding warp function and  scalar field are
    
\begin{equation}
    f(w) = -b \ln \left[2 \cosh( \delta w )\right],
\end{equation}
and

\begin{equation}
    \phi(w)=\sqrt{6b}\arctan\left[ \tanh\left( \frac{\delta w}{2}\right) \right], 
\end{equation}
where $b$ is the AdS-curvature, $b=\sqrt{-\frac{\Lambda}{3}}$, and $\delta$ is brane thickness.  Using the \eqref{eq:F(z)}, we find the class of metamaterials that mimic this kind of brane, they are defined by the tensor components

\begin{equation}\label{eq:Gremm-tensor1}
    \epsilon_{||} = \frac{m^2 + 6bh^2 \arctan^2\left[\tanh\left(\frac{z\delta}{2}\right)\right]}{\left[2 \cosh(z\delta)\right]^{b}},
\end{equation}

\begin{equation}\label{eq:Gremm-tensor2}
    \epsilon_{\perp}=-1.
\end{equation}

We could simplify the expression \eqref{eq:Gremm-tensor1} by  rescaling \eqref{eq:meta-metric} as $ds^2_{an}=m^2d\bar{s}^2_{an}$.  For the same reason, we perform the changes  $z\delta \rightarrow \bar{z}$, $h=m\bar{h}$. As a result, the permittivity components are redefined as
\begin{equation}\label{eq:grem-1}
        \bar{\epsilon}_{||} = \frac{1 + 6b\bar{h}^2 \arctan^2\left[\tanh\left(\frac{\bar{z}}{2}\right)\right]}{\left[2 \cosh(\bar{z})\right]^{b}},
\end{equation}

\begin{equation}
        \bar{\epsilon}_{\perp}=-\frac{1}{m},
\end{equation}
for the metric $d\bar{s}_{an}$. 
This representation means that particles with different masses could be mimicked by changing only  the $\bar{\epsilon}_{\perp}$ permittivity component. Still, it is remarkable to see the consistency of our model for small $\Lambda$, 
\begin{equation}
\bar{\epsilon}_{||}\approx 1 - b\left\lbrace \ln\left[2\cosh(z)\right] - 6\bar{h}^2 \arctan^2 \left[\tanh\left(\frac{\bar{z}}{2}\right)\right]\right\rbrace + \mathcal{O}(b^2),  
\end{equation}
that indicates an almost flat analog geometry as it is expected.

\begin{figure}
    \centering
    \includegraphics[scale = 0.6]{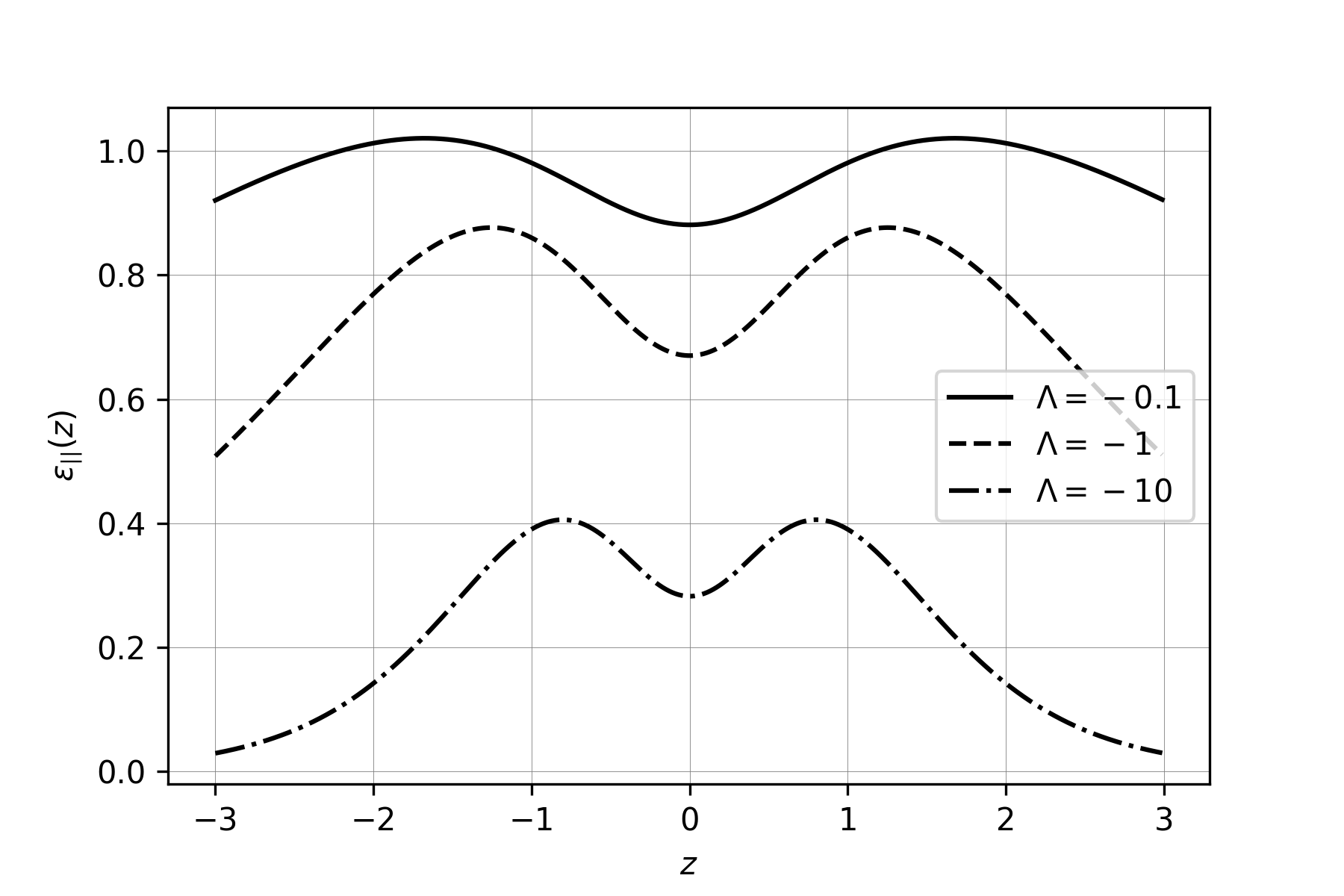}
    \caption{The permittivity $\epsilon_{||}(z)$, \eqref{eq:grem-1}, for  different curvatures in the analog thick brane model, and $h=1$.}
    \label{fig:05}
\end{figure}

\begin{figure}
    \centering
    \includegraphics[scale = 0.6]{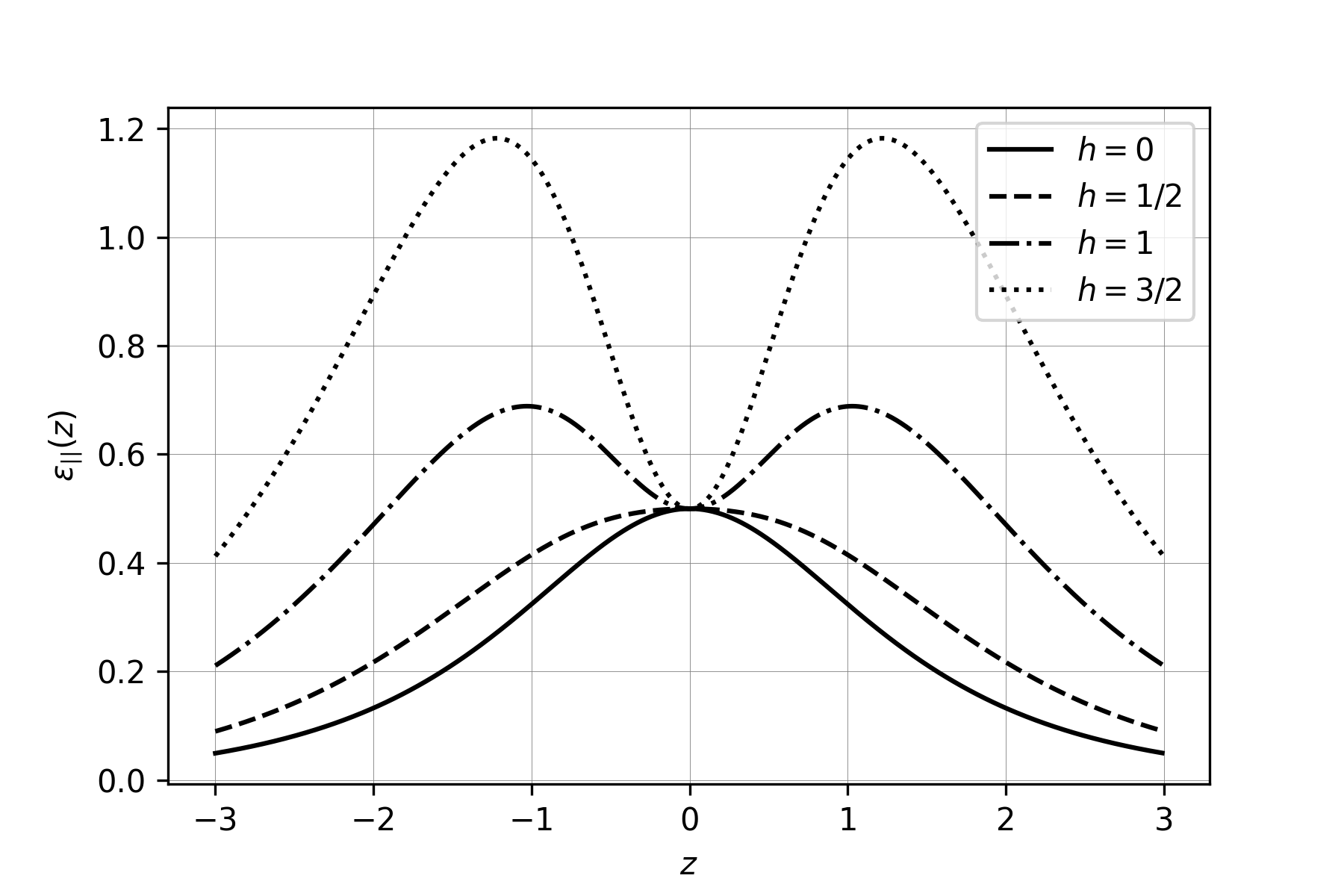}
    \caption{For the thick brane model, we see the $\epsilon_{||}$ plot for different values of the parameter $h$. Here we take
        $\bar{b}=1$ (or $\Lambda=-3$).}
     \label{fig:06}
\end{figure}

In Fig.\eqref{fig:05},   we plot \eqref{eq:Gremm-tensor1} without the bar, where we can see a smoothness of the tensor component in comparison with the previous model, fig.\eqref{fig:01}. This smoothness is also present for different values of analog curvatures and for distinct $h$ values at fig.\eqref{fig:06}.

We can see  qualitatively the behavior of particles in the analog branes using the effective potential description,$V_{eff}$, as defined in \cite{dahia2007confinement}, with the change $f\rightarrow F$. The effective potential becomes
\begin{equation}\label{eq:Veff}
    V_{eff}=e^{2f}\left( 1+\bar{h}^2\phi^2\right)-1.
\end{equation}
In fact, it could reproduce the same behavior as pointed out in the theory, for different interaction factors $h$. If $h=0$, there is no stable confinement, as $h$ increases the analog particles start to oscillate at the center of the analog brane, which represents the confinement.

\section{An analogous asymmetric brane}
Metamaterials can be used to simulate asymmetrical brane models. These warp geometries lose the $Z_2$ symmetry, as proposed, for example, in \cite{bazeia2019new}. They are motivated by cosmological implications \cite{davis2001brane}, by  fermion localization problems \cite{zhao2010fermion}, or even in order to illustrate new braneworld models and its consequences such as gravitational stability \cite{bazeia2019new}.
 
Here we design classes of metamaterials for a particular asymmetrical brane profile \cite{bazeia2014note}, where the scalar field and the warp function are
\begin{eqnarray}
    \phi(w) &=& \tanh{\left(w\right)},\\
    \nonumber \\
    f(w) &=& -\frac{1}{9} \tanh^2{\left(w \right)} + \frac{4}{9}\ln\left[sech(w)\right] - \frac{cw}{3}.
\end{eqnarray}

An important feature of this model is the connection of two distinct asymptotically $AdS_5$ spacetimes, or even the connection of an asymptotically $AdS_5$ with a  $(4+1)$ Minkowski spacetimes. It is important to state that, in metamaterial research,  the junction between two different analog spacetimes had been analysed, as in the case of  $(2+1)$ Minkowski  domain walls \cite{smolyaninov2013minkowski}, or in the case of Klein-Minkowski signature transition \cite{figueiredo2016modeling}.

To understand the asymptotic behavior of this model, it is important to note that the warp factor outside of the brane and the five-dimensional cosmological constants, 
\begin{equation}
    f_{\pm\infty}(w)=-\frac{1}{3}\left(\frac{4}{3}\pm c\right)|w|,
\end{equation}
\begin{equation}
    \Lambda_{\pm}=-\frac{1}{3}\left(\frac{4}{3}\pm c\right)^2.
\end{equation}
It is worth noting that some parameters of the model are taken as one, as the brane thickness, for example. In order to mimic the geodesics in this kind of warp geometry, we need a metamaterial with a permittivity tensor governed by
\begin{gather*}
    \epsilon_{||} = \left[m^2+h^2\tanh^2\left(z\right)\right]e^{-\frac{1}{9}\tanh(z)^2+\frac{4}{9}\ln\left[\sech(z)\right]-\frac{c}{3}z},\\
    \\
     \epsilon_{\perp} = -1.
\end{gather*}
Using $ds^2_{an} = m^2d\bar{s}^2_{an}$ and $h=m\bar{h}$, then
\begin{gather*}
    \epsilon_{||} = \left[1+\bar{h}^2\tanh^2\left(z\right)\right]e^{-\frac{1}{9}\tanh(z)^2+\frac{4}{9}\ln\left[\sech(z)\right]-\frac{c}{3}z},\\
    \\
     \epsilon_{\perp} = -\frac{1}{m}.
\end{gather*}

\begin{figure}[!h]
    \centering
    \includegraphics[scale = 0.6]{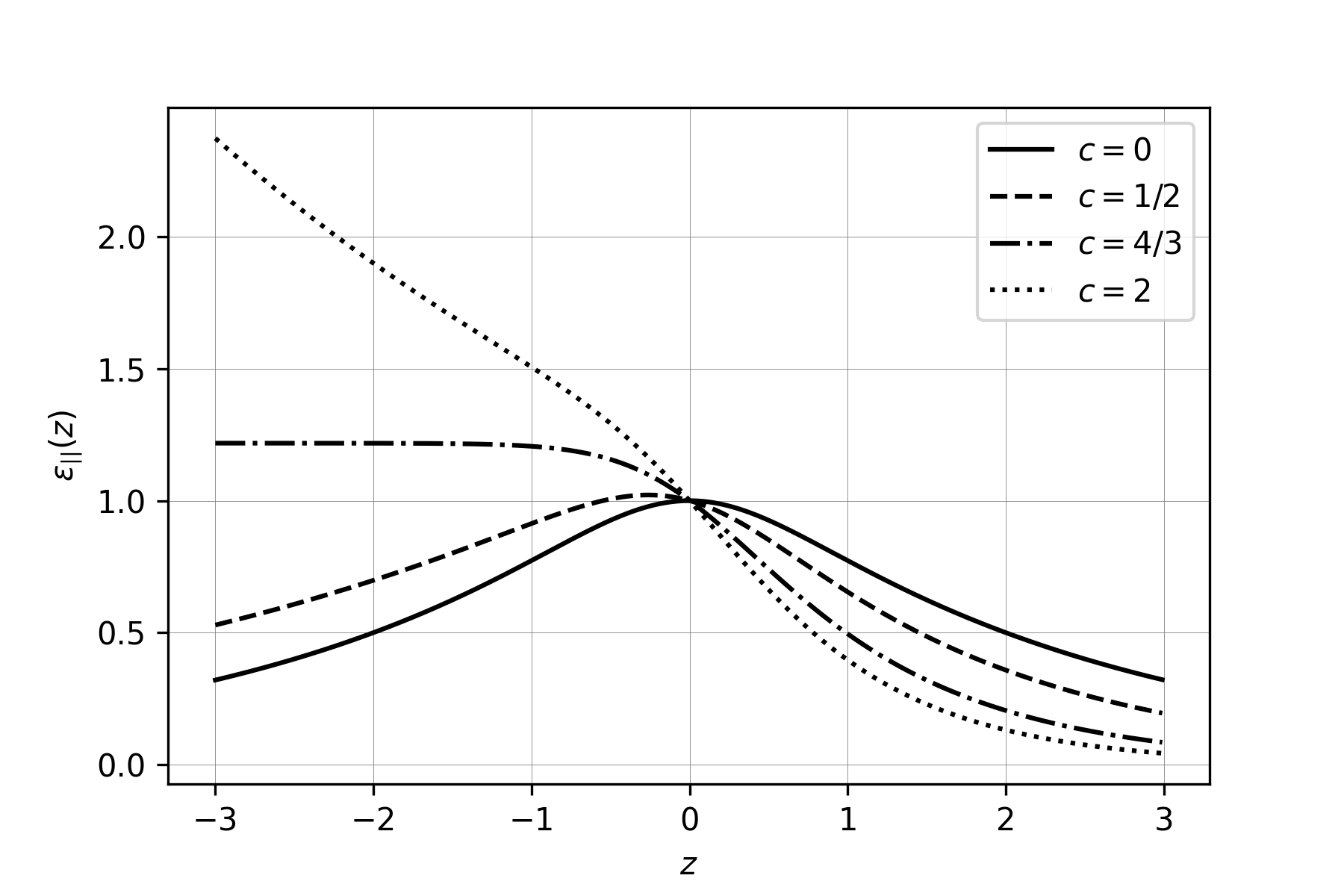}
    \caption{Graphs for $h = 0$. We see that for $c=0$, the permittivity component has a specular symmetry. For $c \neq 0$, the component is asymmetric, and for $c = 2$, it seems to increase indefinitely when $z\rightarrow -\infty$.}
    \label{fig:08}
\end{figure}

\begin{figure}[!h]
    \centering
    \includegraphics[scale = 0.6]{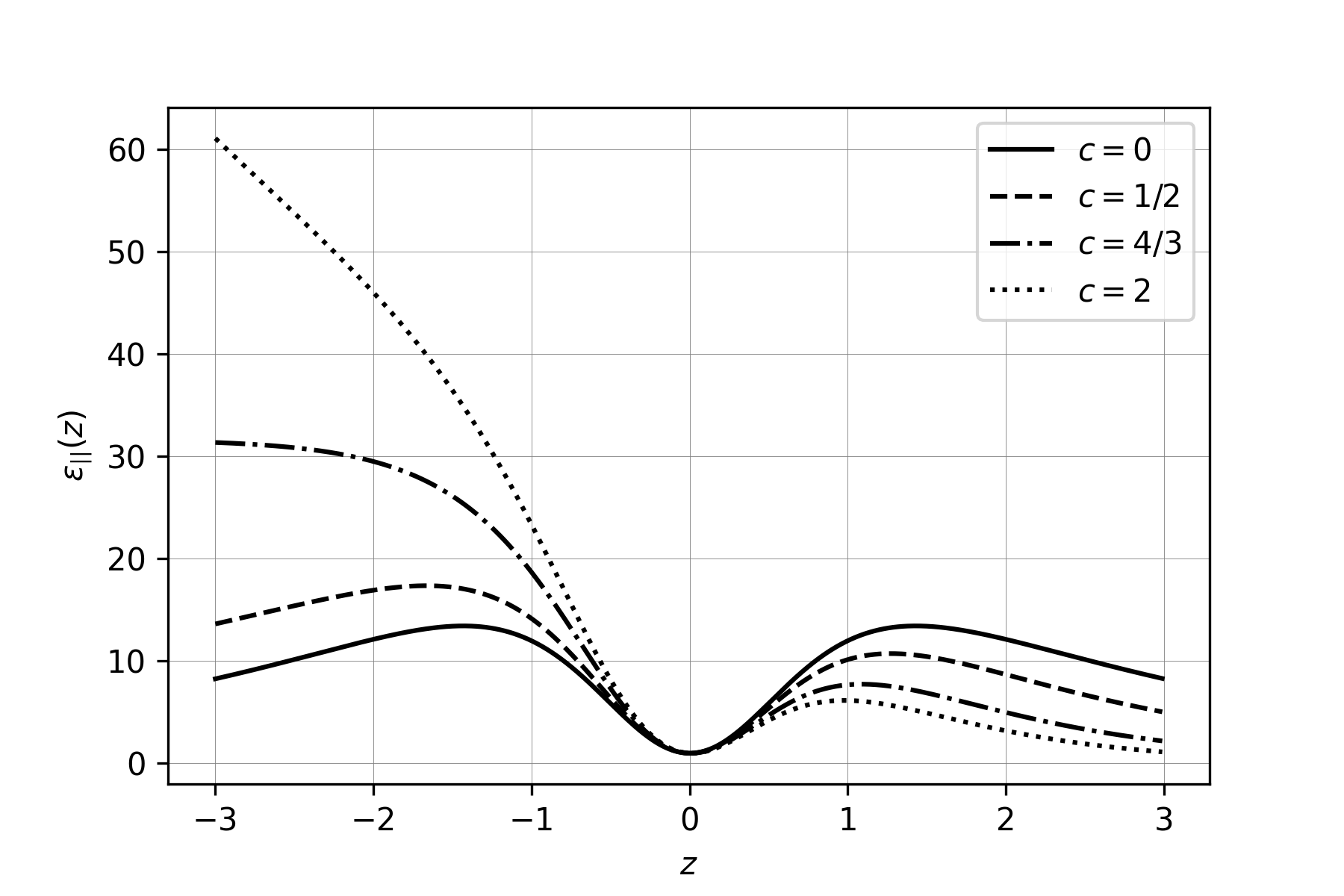}
    \caption{Graphs for $h = 5$. We see an inflection point at the origin, and for $c = 2$, it seems to increase indefinitely when $z\rightarrow -\infty$.}
    \label{fig:09}
\end{figure}

These are the simplified permittivity tensor components. Since the geodesics of $ds^2$ and $d\bar{s}^2$ are equivalent, we use the bar metric and we shall write $\bar{h}$ as $h$ henceforward. In fig. \eqref{fig:08}, we can see distinct types of $\epsilon_{||}$ behaviors for $h=0$. If $c = 0$, it corresponds to a symmetric medium. In the other cases, then, we have asymmetric metamaterials. In fig. \eqref{fig:09}, we see the permittivity components that mimic the Yukawa-like interaction, and also that $c$ breaks the symmetry of the medium.
\subsection{Analog massive particle transverse motion}
Here we intend to describe the behavior of analog particles following
the analog extra dimension, the $z-$direction.  The geodesic equation for $z(\lambda)$ is $\ddot{z} + F'(z)(1+\dot{z}^2) = 0$,
which we can reduce to an autonomous planar dynamical system as follow 
\begin{eqnarray}
    \dot{z} &=& p\label{zp}\\
    \nonumber\\
    \dot{p} &=& -F'(z)(1+p^2)\label{pp}
\end{eqnarray}
where,
\begin{equation}
    F'(z) = \tanh(z)\left[\frac{h^2\sech^2(z)}{1+h^2\tanh^2(z)}-\frac{2}{9}\left(\sech^2(z)+2\right)\right]-\frac{c}{3}. \label{eq:critical}
\end{equation}
From \eqref{zp} and \eqref{pp}, we can find the critical (equilibrium) points by solving the equations $\dot{z} = \dot{p} = 0$, which implies that all critical points lie on the axis $p=0$ and are located at the roots of \eqref{eq:critical}. 

In other words, the set of critical points, $\{z_i\}$, are given by the solutions for $F'(z_i) = 0$. Therefore, let us analyze some cases of interest determined by the values ascribed to the  parameters $c$ and $h$.
\subsubsection{Case $c = 0$: Analogue $AdS_5$ bulk}
This model was originally proposed in \cite{kehagias2001localized}, which has the feature of localized massless chiral fermions. In this case, we still have the $Z_2$ symmetry, since the asymmetry parameter $c$ is null.  The corresponding metamaterials, engineered taking $c=0$, can realize an asymptotic five-dimensional Anti-de Sitter spacetime, with $\Lambda_{\pm} = -\frac{16}{27}$, in three dimensions.

If $h = 0$, then there is only one critical point at the origin of phase space. As already mentioned, the physical role of the parameter $h$ is associated with the coupling with the scalar field of the model. Therefore, as this parameter increases, increasing the coupling (or the thickness of the brane), we have the appearance of a region of stability around the origin at the phase space. In this stability region, which grows with the value of $h$, the solutions are represented by orbits in phase space, evidencing a periodic behavior that must be interpreted as mimicking the particle confinement in the brane.

\begin{figure}[h]
    \centering
    \includegraphics[width = 0.5\textwidth]{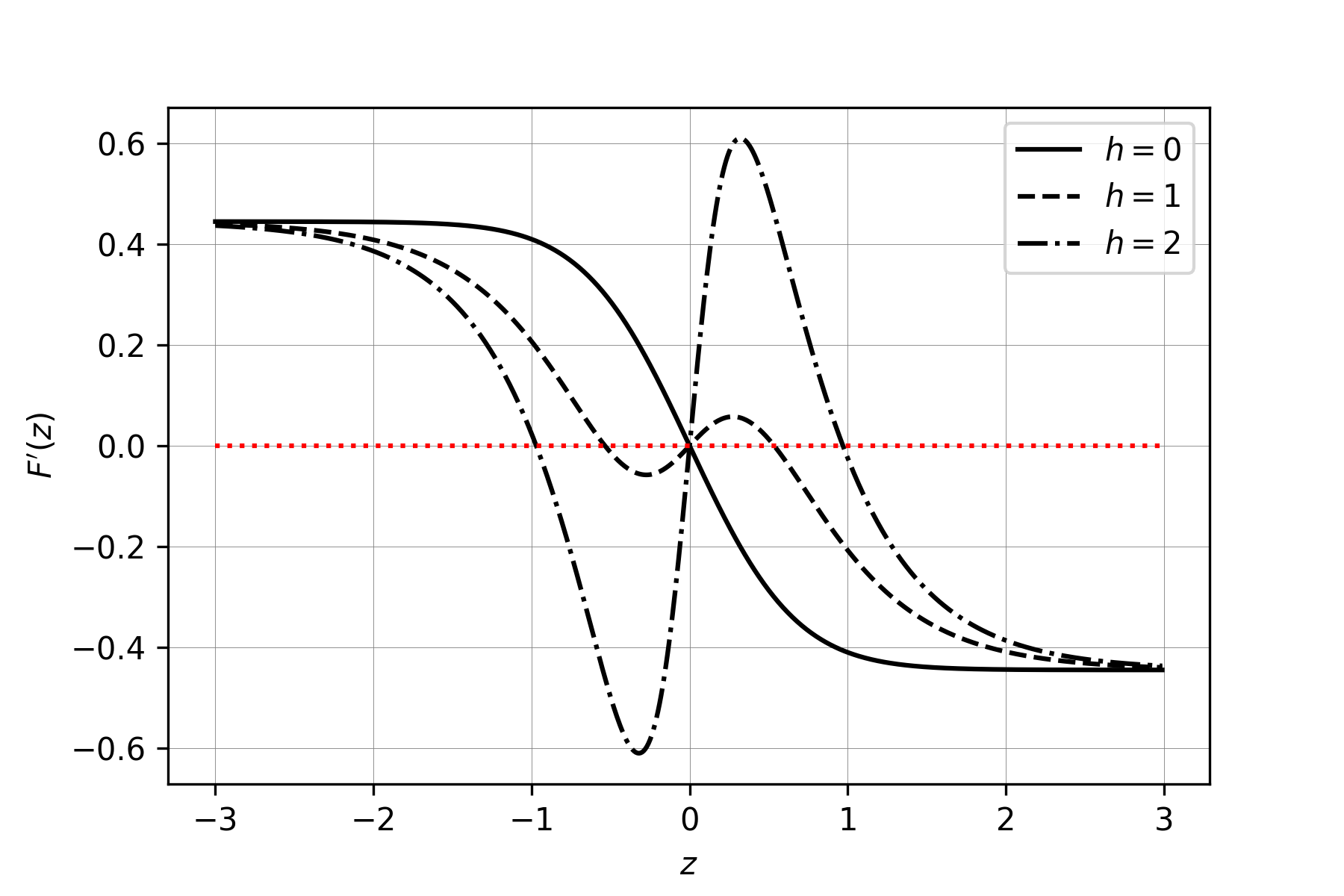}
    \caption{Case $c = 0$. For $h=0$, there is no critical points. For $h = \lbrace 1, 2\rbrace$, there are three critical points.}
    \label{fig:c0hs}
\end{figure}

However, when assigning non-zero values to $h$, other roots arise for \eqref{eq:critical} developing other critical points in the vicinity of the origin. The behavior of \eqref{eq:critical} for $h = \lbrace 0, 1, 2\rbrace$ is plotted in Fig.\eqref{fig:c0hs}.

\begin{figure}[h]
    \centering
    \includegraphics[width = 0.4\textwidth]{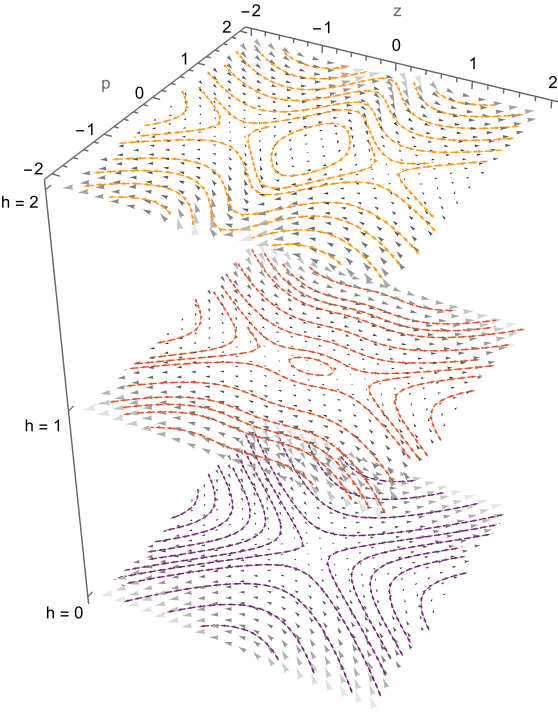}
    \caption{Case $c = 0$. Phase portraits of the system given in \eqref{zp} and \eqref{pp} for $h = \lbrace 0, 1, 2\rbrace$. In the cases for $h={1,2}$, we see that the orbits resemble the simple pendulum oscillation phase portrait orbits for non-linear regime.}
    \label{fig:c0Evolution}
\end{figure}

In Fig. \eqref{fig:c0Evolution} we gather the phase portraits for each value of $h$, showing how the behavior of the critical points changes in the case $c = 0$ as a function of $h$. For $h=0$, the analogue particles are reflected from the brane center,$z=0$. For $h\neq 0$ values the particles  are reflected near the saddle points. Also, for $h\neq 0$, the confinement mechanics make particles oscillate inside the brane, $h=1$, or oscillate crossing the brane, $h=2$. 

\subsubsection{Case $c = 1/2$: Analogue Asymmetric $AdS_5$ bulk}
\begin{figure}[h]
    \centering
    \includegraphics[width = 0.4\textwidth]{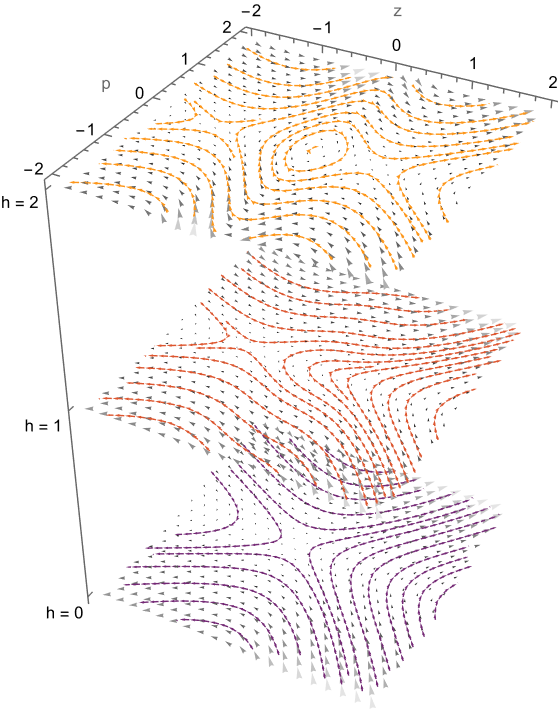}
    \caption{Case $c = 1/2$. Phase portraits of the system given in \eqref{zp} and \eqref{pp} for $h = \lbrace 0, 1, 2\rbrace$. Comparing to fig. \eqref{fig:c0Evolution}, we see a dislocations of the critical points.}
    \label{fig:c12Evolution}
\end{figure}

In metamaterials engineered taking $c=1/2$, there is a manifestation of the asymmetry of the model. Here, two different asymptotically Anti-de Sitter spacetimes are mimicked, with an analogous $\Lambda_{-}=-\frac{25}{108}$ as $z\rightarrow -\infty$ and
$\Lambda_{+}=-\frac{121}{108}$  as $z\rightarrow +\infty$.
In Fig.  \eqref{fig:c12Evolution}, we can see the behavior of the geodesics in a phase plot for different $h$ parameters.

\begin{figure}[h]
    \centering
    \includegraphics[width = 0.5\textwidth]{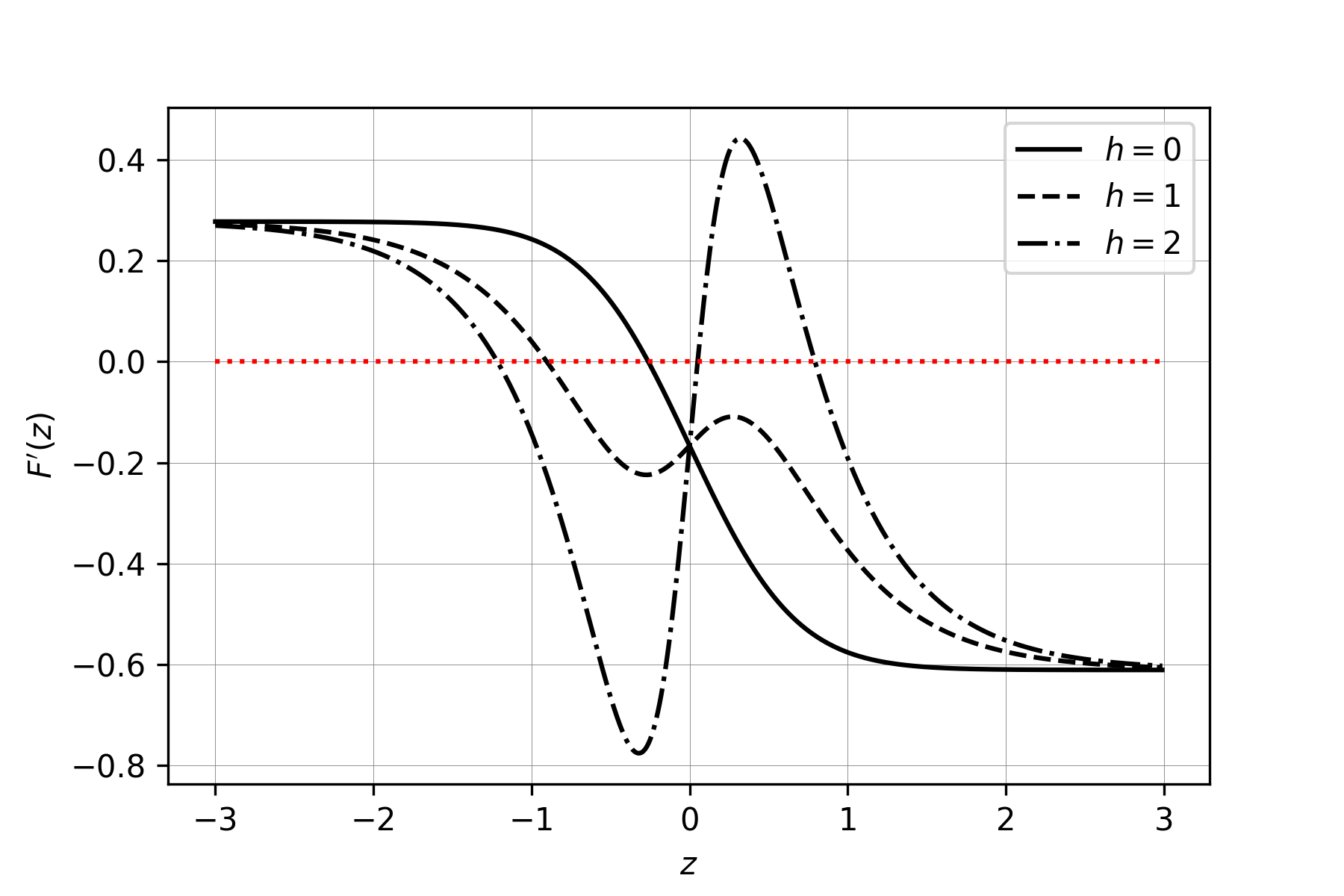}
    \caption{Case $c = 1/2$ and $h = \lbrace 0, 1, 2\rbrace$. The roots represent the critical points.}
    \label{fig:c12hs}
\end{figure}

For $h=0$, there is only one  critical point, a saddle point, at the left side, the analog particles are repelled from it. It is also possible,  for some restricted conditions, that at this point the particles could be at rest, but this equilibrium is unstable. 
For $h=1$, there is still one saddle point. For $h= 2$, two new equilibrium points appear, the stable one is a center near the origin. The number and exact position of the critical points, the roots of $F'$, are shown in \eqref{fig:c12hs}. 
 
 Furthermore, if we write an effective potential for this case, following \eqref{eq:Veff}, it can be deduced that the analog particles need less energy to go to the right than to go to the left in for all these $h$ values. This is a result of the asymmetrical analog bulk.
\subsubsection{Case $c = 4/3$: An analogue $\mathbb{M}_5$-$AdS_5$ bulk}
\begin{figure}[h]
    \centering
    \includegraphics[width = 0.5\textwidth]{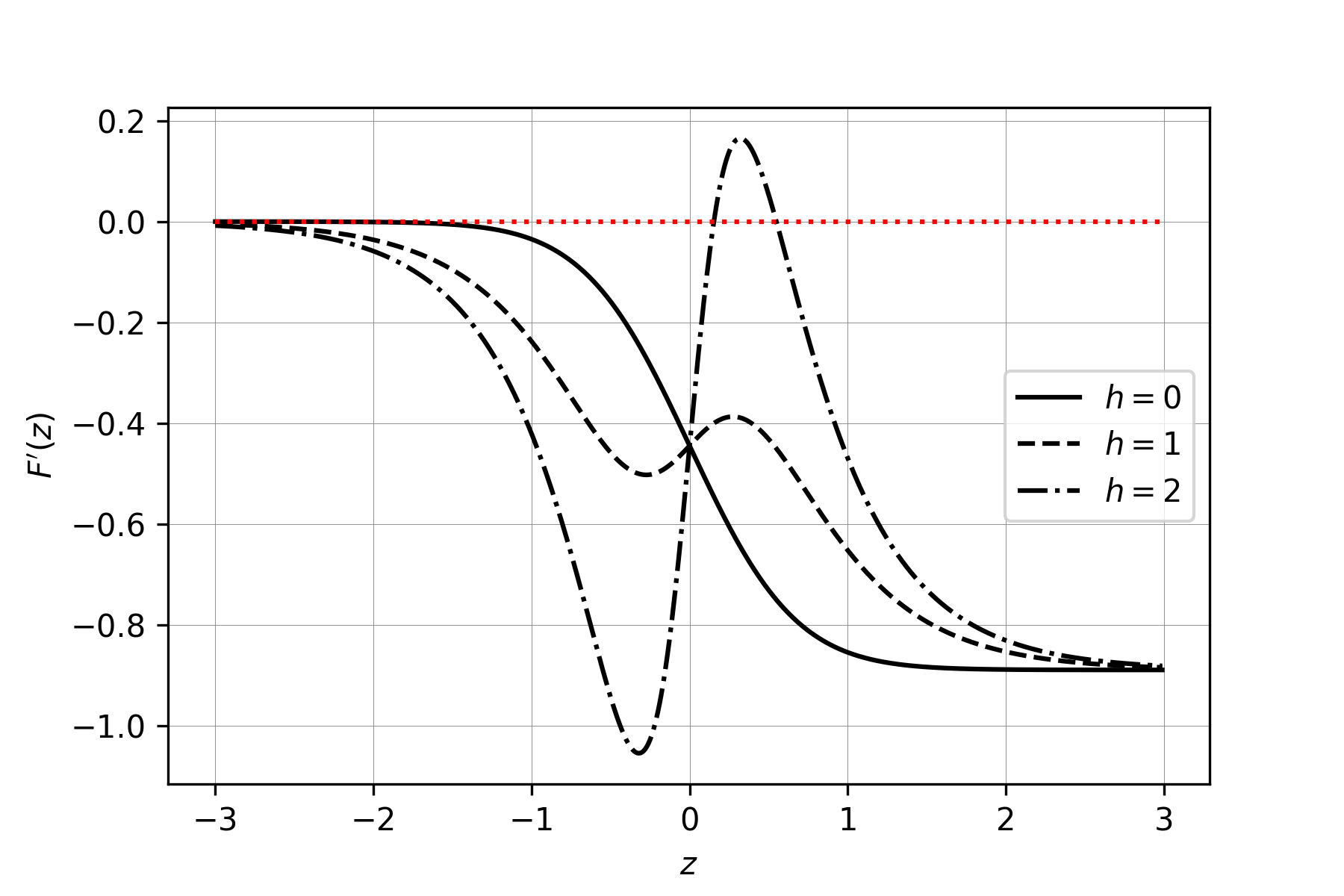}
    \caption{Case $c = 4/3$. The critical points are asymptotic for
    $h=\{0,1\}$.}
    \label{fig:c43hs}
\end{figure}

\begin{figure}[h]
    \centering
    \includegraphics[width = 0.4\textwidth]{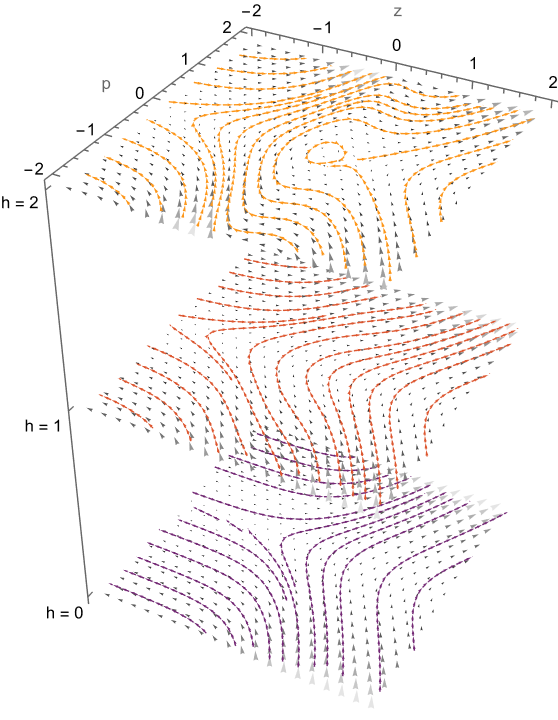}
    \caption{Case $c = 4/3$. Phase portraits of the system given in \eqref{zp} and \eqref{pp} for $h = \lbrace 0, 1, 2\rbrace$.The left trajectories tends to straight lines since asymptotically there is no gravity.}
    \label{fig:c43Evolution}
\end{figure}

Metamaterials designed with $c=\frac{4}{3}$  connect an analog $(4+1)$ Minkowski space-time  to an analogue $AdS_5$ space-time, see Fig. \ref{fig:c43hs} for a few $h$ values. As we can see from $F'(z)=0$  and also $F' \longrightarrow 0$ as $z\longrightarrow -\infty$, this indicates that the analogous  $\Lambda_{-}$ on the left side is null. From Fig.  \ref{fig:c43Evolution}, it is possible to  note that we only have critical points at $h=2$, one center, and one saddle point, both on the right side. This means that analog particles could be trapped in a tiny region between both of them. Independent of the values, $h=\{0,1,2\}$, the light has a tendency to go to the right hand side direction, because they are repelled from the critical point.


\begin{figure}[h]
    \centering
    \includegraphics[width = 0.5\textwidth]{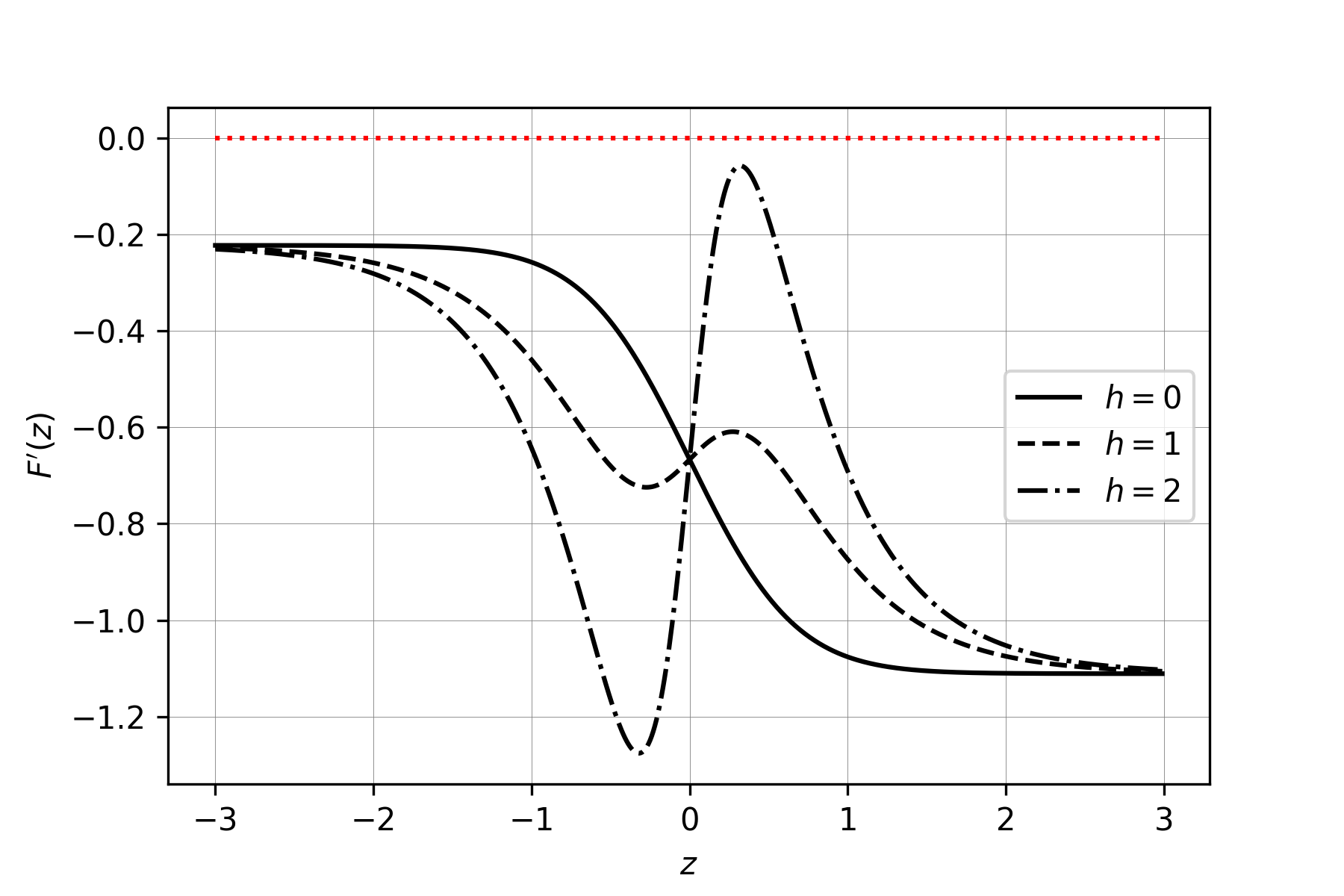}
    \caption{Case $c = 2$ and $h = \lbrace 0, 1, 2\rbrace$. We see that there is no critical points.}
    \label{fig:c2hs}
\end{figure}
\begin{figure}[h]
    \centering
    \includegraphics[width = 0.4\textwidth]{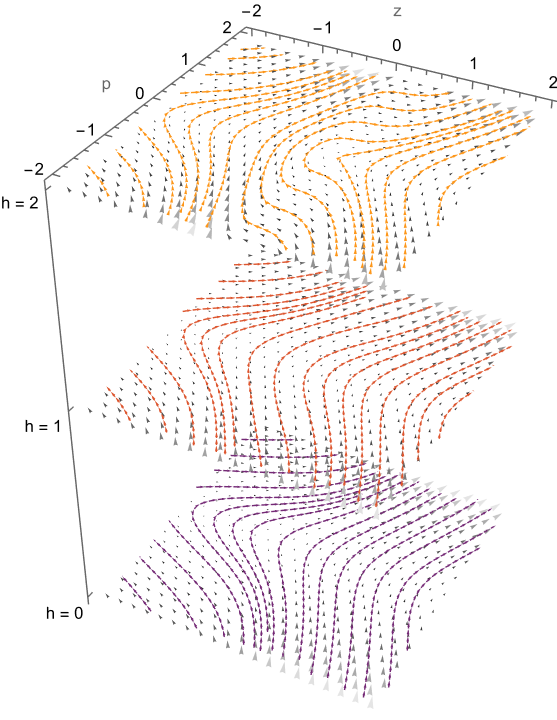}
    \caption{Case $c = 2$. Phase portraits of the system given in \eqref{zp} and \eqref{pp} for $h = \lbrace 0, 1, 2\rbrace$. We see a tendency that particles are  repleted to the right.}
    \label{fig:c2Evolution}
\end{figure}

\subsubsection{Case $c = 2$: An analogue divergent brane}
Metamaterials with $c = 2$ can simulate a brane model connecting an $AdS_5$ with $\Lambda_{-}=-\frac{4}{27}$ to another one with $\Lambda_{+}=-\frac{100}{27}$. We can see see that at the left side, the warp factor asymptotically diverges.
From the figs. \ref{fig:c2hs} and \ref{fig:c2Evolution}, there is no critical points. Also, we can expect that analog particles at the center should naturally go to the  right. There is confinement for higher values of $h$, as $h\geq 5$ (see the Appendix).


\subsubsection{Summary of plots}
In summary, we point out some general features of all the phase diagrams here.  If $c\neq 0$, then there is no critical point at the origin as we show in the Appendix. In fact, $\dot{p}=-\frac{c}{3}$, such that $c$ determines the ``repulsion'' at the brane center $(z=0)$. Particles at rest also moves according to $\dot{p}=-\frac{c}{3}$, independent of their positions. 
Another general feature determined by $c$ is the asymptotic behavior, as $z \rightarrow \pm\infty$ then $p(\lambda)\rightarrow \tan\left((\frac{c}{3}\mp\frac{4}{9})\lambda\right)$.
In this brane description, we use  positive values for $c$, however, it is possible to construct analog warp geometry models for negative $c$ values.

\section{Conclusion\label{sec:Concl}}

In this article, we show profiles of materials that can mimic the geodesics  of massive particles in some braneworld models, with and without the confinement mechanism of bosonic particles. It is known that there are other mechanisms of confinement in the literature \cite{souza2021confinement}, however, it requires more study in order to design metamaterials for them.

In metamaterials modeled with \eqref{eq:F(z)}, in the case of analogue thick branes, the light ray particles get an associated analog mass that depends on the metamaterial parameters. Then, particles with different masses can not be represented by the same medium. As a consequence, we cannot simulate the split particles in different slices of the brane as it could occur for fermions according to  \cite{dahia2013electric}. 
    
In order to map a photon, we could use a hyperbolic liquid crystal metamaterial. Given the line element $ds_{an}^2=\epsilon_{\perp}dr^2-|\epsilon_{||}|\left(r^2d\phi^2 +dz^2\right)$, where we associate the $r$ coordinate with the extra-dimension, and $z$ and $\phi$ to coordinates in the  brane. Choosing $\epsilon_{\perp}=1$ and
$\epsilon_{||}=e^{-f(r)}$. We easily get that massless particles move according to the equations $\ddot{r}+f'\dot{r}^2 -e^{f}r\dot{\phi}^2=0$ and $\dot{\phi}=\frac{L_\phi}{r^2e^{f}}$, where $L_\phi $ is a constant of the motion. Therefore, the photon motion in the $r-$direction is given by $\ddot{r}+f'\dot{r}^2-\frac{e^{-f}}{r^3}L_{\phi}=0$ and for radial emission, $L_{\phi}=0$, the equation gets the form $\ddot{r}+f'\dot{r}^2 =0$, that simulates the transverse  motion along the extra-dimension, according to \eqref{eq:photon}. 

Concerning the asymmetrical analog brane, we find that light could  stop at the saddle points. This seems to have a connection with slow light phenomena, however according to \cite{reza2008can} this should not happen, and as a consequence, we need the electromagnetic wave approach to understand better this issue. Still, we can infer general features for our models from our results. Since the metric \eqref{eq:meta-metric}is valid and $\epsilon_{\perp}$ is constant, then $\frac{d^2\epsilon_{||}}{dz^2}>0$ at a critical point, we have that the confinement is possible, however, if $\frac{d^2\epsilon_{||}}{dz^2}<0$ at the critical point, then there is no confinement of light at this region.  

Finally, it seems to us, that our  proposals could be adapted to study the geodesic motion in bi-dimensional sub-manifolds of six-dimensional space-time warp geometries such as the Gergheta-Shaposhnikov  or string-cigar models \cite{gherghetta2000localizing,silva2012gravity, costa2015gauge}, where the two new co-dimensions are $r$ and $\theta$. 
Furthermore, since there are few results in the extra-dimensional analog models, for example \cite{barcelo2003braneworld}, we expect  to shed light on  Randall-Sundrum like models at the lab.

\acknowledgments{We would like to thank C. Romero and F. Dahia for their helpful discussions. This work was supported by Conselho Nacional de Desenvolvimento Cient\'{i}fico e Tecnol\'{o}gico (CNPq), grants Nos. 150480/2021-0 (ABB) and 311798/2021-7 (FM).}

\bibliography{sample}

\appendix

\section{Phase portraits}
Here we present some additional phase portraits  of the dynamical system defined by Eqs. \eqref{zp} and \eqref{pp} of the analog asymmetrical branes  (section $IV$), for some values of $c$ and $h$. Where they appear, the red dots represent saddle points (unstable), and the green dots, centers (stable).
\begin{figure}[H]
    \centering
    \includegraphics[width = 0.4\textwidth]{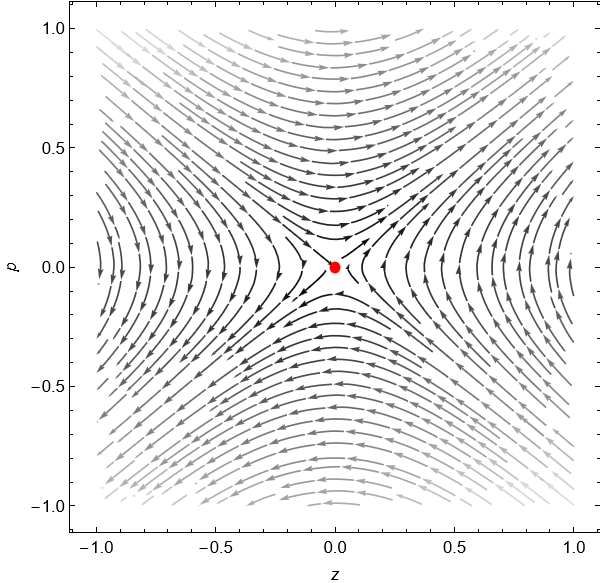}
    \caption{The case $c = 0$ and $h = 0$. }
    \label{fig:c0h0}
\end{figure}

\begin{figure}[H]
    \centering
    \includegraphics[width = 0.4\textwidth]{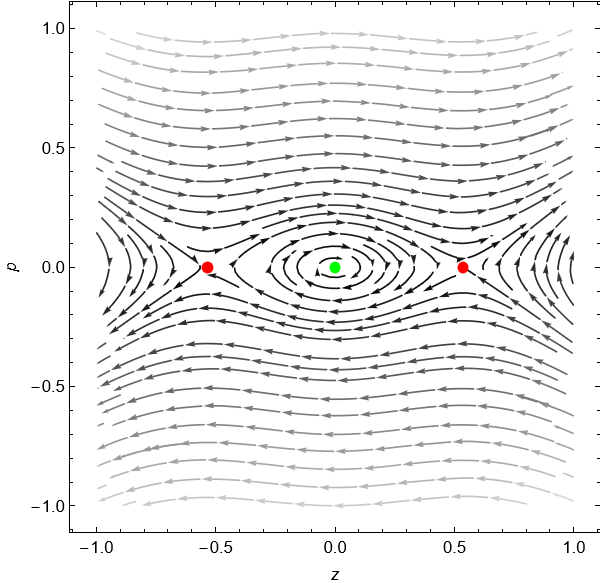}
    \caption{The case $c = 0$ and $h = 1$. }
    \label{fig:c0h1}
\end{figure}

\begin{figure}[H]
    \centering
    \includegraphics[width = 0.4\textwidth]{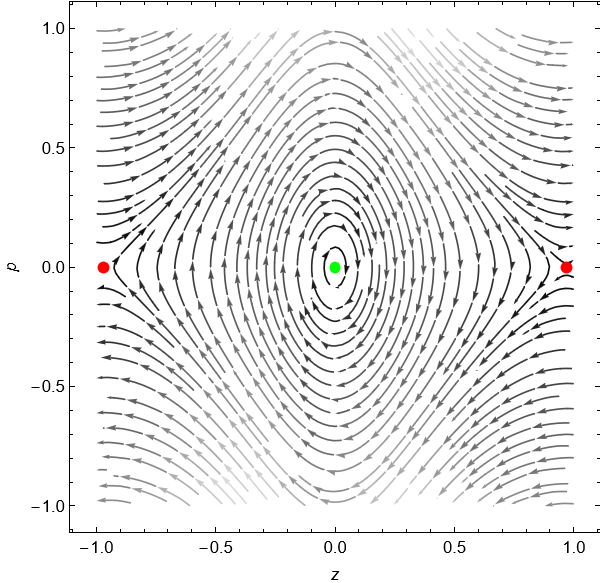}
    \caption{The case $c = 0$ and $h = 2$. }
    \label{fig:c0h2}
\end{figure}

\begin{figure}[H]
    \centering
    \includegraphics[width = 0.4\textwidth]{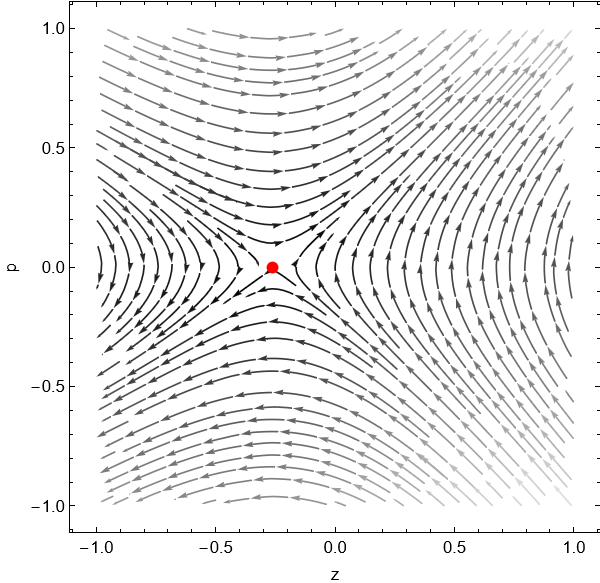}
    \caption{The case $c = 1/2$ and $h = 0$. }
    \label{fig:c12h0}
\end{figure}

\begin{figure}[H]
    \centering
    \includegraphics[width = 0.4\textwidth]{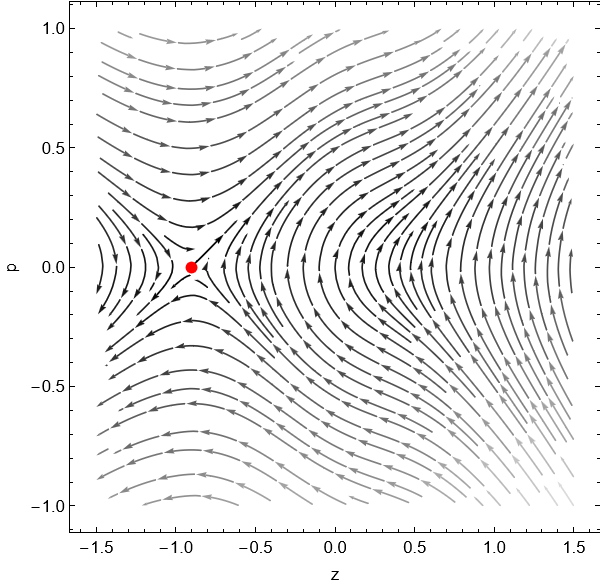}
    \caption{The case $c = 1/2$ and $h = 1$. }
    \label{fig:c12h1}
\end{figure}

\begin{figure}[H]
    \centering
    \includegraphics[width = 0.4\textwidth]{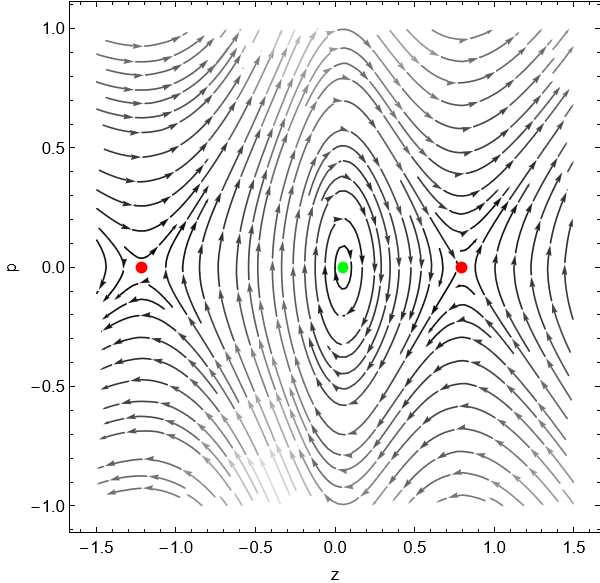}
    \caption{The case $c = 1/2$ and $h = 2$.}
    \label{fig:c12h2}
\end{figure}

\begin{figure}[H]
    \centering
    \includegraphics[width = 0.4\textwidth]{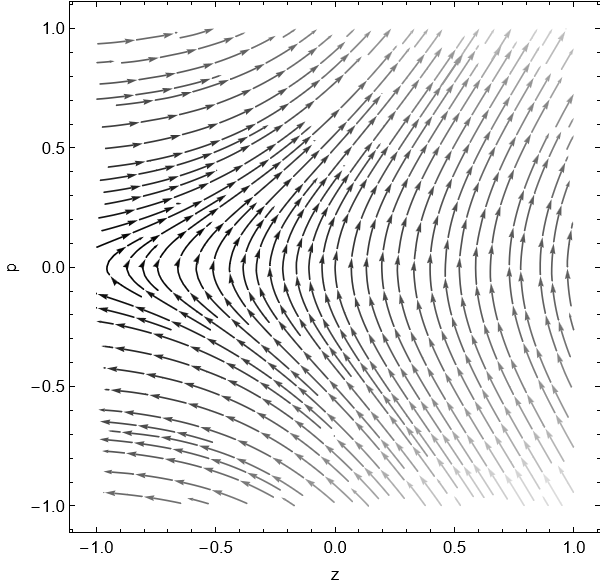}
    \caption{The case $c = 4/3$ and $h = 0$. In this case, there are no critical points.}
    \label{fig:c43h0}
\end{figure}

\begin{figure}[H]
    \centering
    \includegraphics[width = 0.4\textwidth]{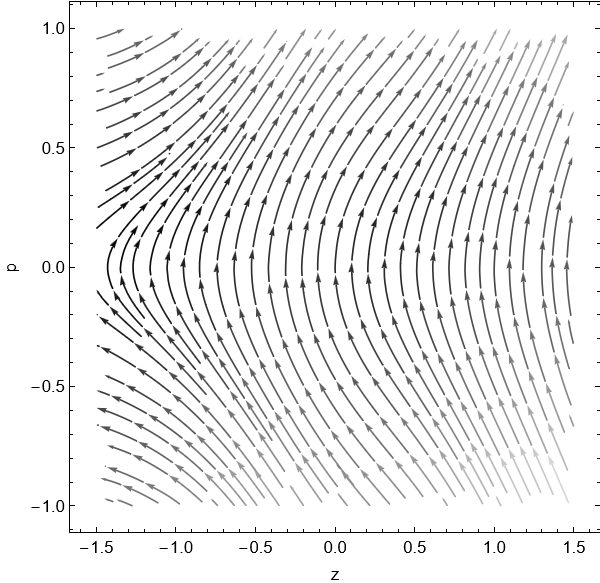}
    \caption{The case $c = 4/3$ and $h = 1$.   In this case, there are no critical points.}
    \label{fig:c43h1}
\end{figure}

\begin{figure}[H]
    \centering
    \includegraphics[width = 0.4\textwidth]{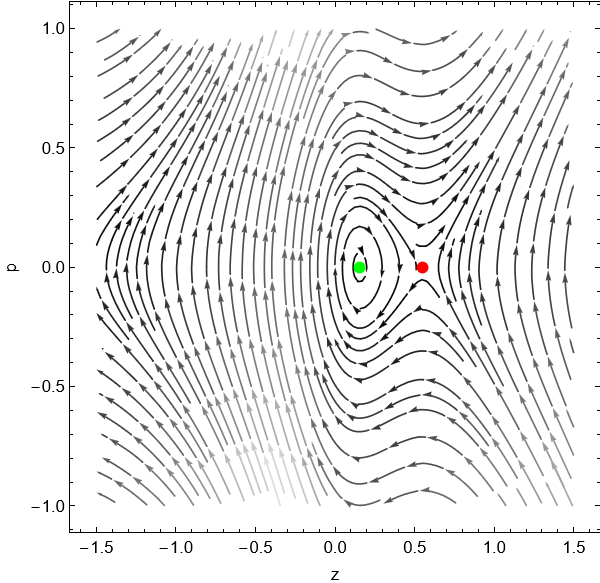}
    \caption{The case $c = 4/3$ and $h = 2$.  }
    \label{fig:c43h2}
\end{figure}

\begin{figure}[H]
    \centering
    \includegraphics[width = 0.4\textwidth]{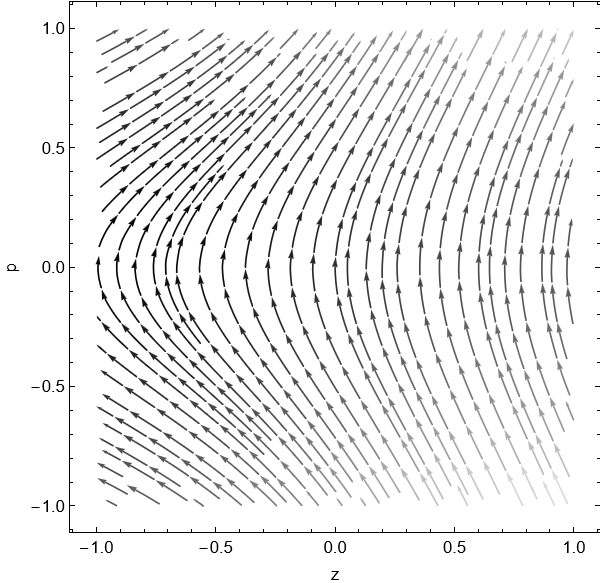}
    \caption{The case $c = 2$ and $h = 0$.  In this case, there are no critical points.}
    \label{fig:c2h0}
\end{figure}

\begin{figure}[H]
    \centering
    \includegraphics[width = 0.4\textwidth]{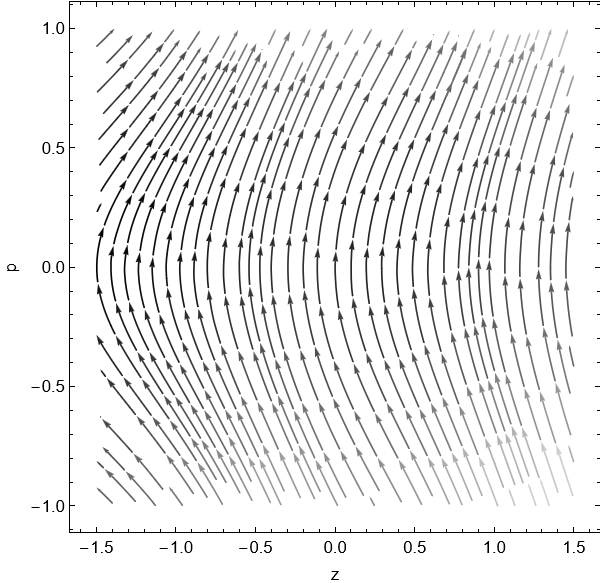}
    \caption{The case $c = 2$ and $h = 1$.   In this case, there are no critical points.}
    \label{fig:c2h1}
\end{figure}

\begin{figure}[H]
    \centering
    \includegraphics[width = 0.4\textwidth]{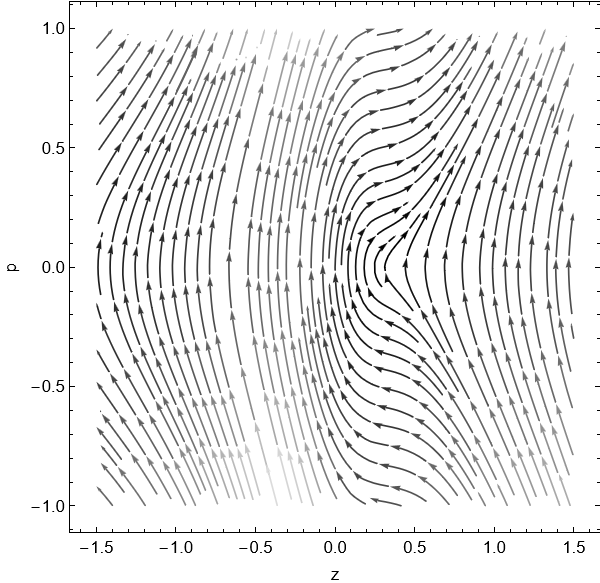}
    \caption{the case $c = 1/2$ and $h = 2$.  In this case, there are no critical points.}
    \label{fig:c2h2}
\end{figure}

\begin{figure}[H]
    \centering
    \includegraphics[width = 0.4\textwidth]{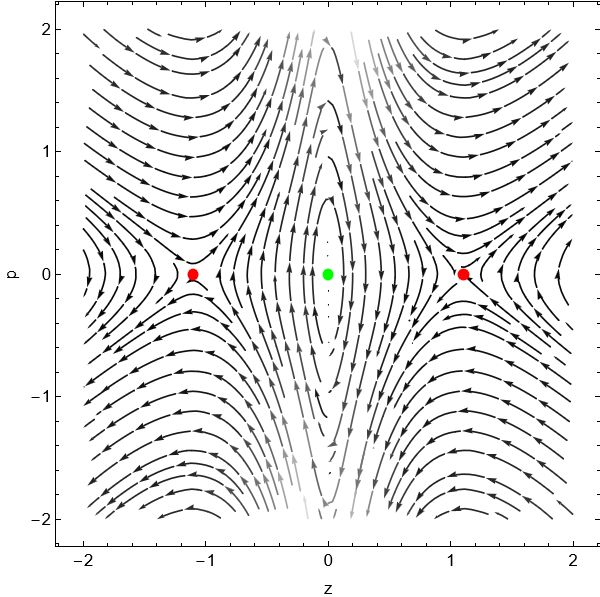}
    \caption{The case $c = 0$ and $h = 5$.  }
    \label{fig:h5c0}
\end{figure}

\begin{figure}[H]
    \centering
    \includegraphics[width = 0.4\textwidth]{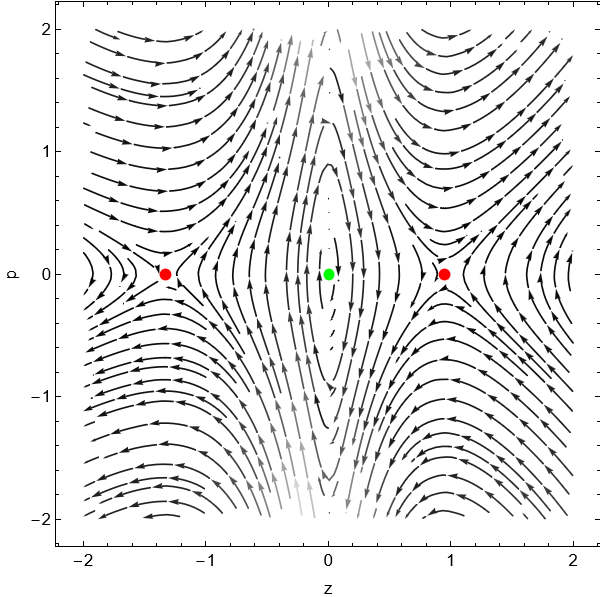}
    \caption{The case $c = 1/2$ and $h = 5$.  }
    \label{fig:h5c12}
\end{figure}

\begin{figure}[H]
    \centering
    \includegraphics[width = 0.4\textwidth]{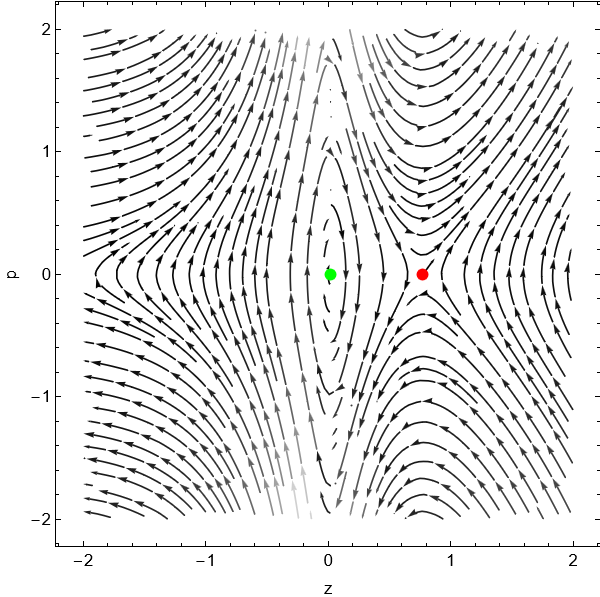}
    \caption{The case $c = 4/3$ and $h = 5$.}
    \label{fig:h5c43}
\end{figure}

\begin{figure}[H]
    \centering
    \includegraphics[width = 0.4\textwidth]{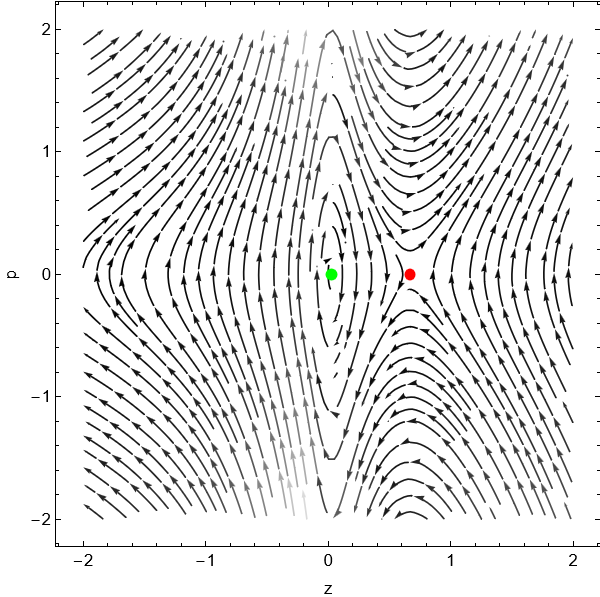}
    \caption{The case $c = 2$ and $h = 5$. }
    \label{fig:h5c2}
\end{figure}

\end{document}